\documentstyle[prl,aps,array,epsf,twocolumn,psfig]{revtex}
\def\be{\begin{equation}}
\def\ee{\end{equation}}
\def\ba{\begin{array}}
\def\ea{\end{array}}

\begin{document}
\twocolumn[\hsize\textwidth\columnwidth\hsize\csname
 @twocolumnfalse\endcsname      

\title{Bubbling and Large-Scale Structures in Avalanche Dynamics}

\author{Supriya Krishnamurthy$^{a}$, Vittorio Loreto$^{a}$ and
 St\'ephane Roux$^{b}$}

\pagestyle{myheadings}
\address{
$a)$ Laboratoire Physique et M\'ecanique des Milieux
H\'et\'erogenes,\\
Ecole Sup\'erieure de Physique et Chimie Industrielles,\\
10 rue Vauquelin, 75231 Paris cedex 05, France.\\
$b)$  Surface du Verre et Interfaces,\\ 
Unit\'e Mixte de Recherche CNRS/Saint-Gobain,\\
39 Quai Lucien Lefranc, 93303 Aubervilliers Cedex, France.\\}  

\maketitle
\date{\today}
\begin{abstract}
Using a simple lattice model for granular media, we present
a scenario of self-organization that we term self-organized
structuring where the steady state has several unusual features:
(1) large scale space and/or time inhomogeneities and (2)
the occurrence of a non-trivial peaked distribution of large events
which propagate like ``bubbles'' and have  a well-defined frequency 
of occurrence. We discuss the applicability of such a scenario for other 
models introduced in the framework of self-organized criticality.

{ {\bf PACS numbers}:  45.70.-n,74.80.-g,05.65.+b}
\end{abstract}
\smallskip
\vskip2pc]


One of the major challenges of statistical physics in recent 
years has been to address the question of the ubiquity of fractality and 
scale invariance phenomena in Nature\cite{Mandelbrot}. The occurrence of
such scaling properties in an extremely broad (and unrelated) class of 
problems, calls for a concept of extreme generality,
beyond the level of
proposing a model suited for any specific phenomenon.
A major
step in this direction was taken by Bak {\it et al}\cite{BakTan} 
who proposed the notion of self-organized critical phenomena (SOC).

The philosophy of SOCis to note that a 
precise dynamics can be associated with most second order critical 
phenomena. This corresponds generically to an infinitesimal external 
forcing of the system which renders the critical point an attractor 
of the dynamics.
Therefore, the natural evolution of the system drives it to the vicinity
of a critical point, and the steady state displays scaling properties
which can in some instances be related to the ''static'' critical
properties of the phase transition at the critical point.
Qualitative predictions of this scenario are thus the appearance of
power-law distributions of avalanches, $1/f$ noise, and more generally the
absence of characteristic length or time scales in the dynamics. 
Examples range from sand-piles\cite{BakTan} to earthquakes
\cite{CarLan,CarLan2}, from stock market fluctuations\cite{soc-economy} 
to the speciation of living organisms on Earth\cite{SneBak}.

The purpose of the present letter is to report on some unusual
features of self-organization occurring during the restructuring of a
granular medium \cite{grain}
under internal avalanches \cite{KriLor,Ball}. In
particular we show that the system spontaneously breaks 
the spatial homogeneity to develop large-scale structures
which  trigger (almost) periodic emission of large
avalanches that propagate like solitary bubbles. As a consequence, the
avalanche distribution exhibits a non-trivial bump for large
avalanche sizes. This mechanism can  
coexist with small scale scaling in some cases while controlling it
in other cases. We term this phenomenon
self-organized structuring.  We show that this
mechanism sheds a new light on phenomena usually interpreted in the
framework of SOC, such as the celebrated Burridge-Knopoff model
\cite{BurKno} for solid friction and earthquakes\cite{Sho}.  It also
calls attention to finite size effects, which may alter the proper
understanding of the large scale physics at play.  The self-organized
structuring presented here may also help to understand the often
debated question of the relevance of the original model of SOC, the
sand-pile model, to real avalanches in granular media as observed
experimentally\cite{oslo}.



We illustrate self-organized structuring using a model for describing
the restructuring of a granular medium caused by an infinitesimal
perturbation (a quasi-static flow of particles out of a vessel).  The
description of the granular medium is based on the ``Tetris'' model
\cite{tetris}, a toy model designed to account for geometric 
frustration.  This model has been shown to reproduce a number of complex 
features observed experimentally in granular matter, such as the 
slow dynamics of compaction under vibration\cite{Knight}, 
segregation\cite{segtet} etc.

\begin{figure}
\bigskip
\centerline{
      \psfig{figure=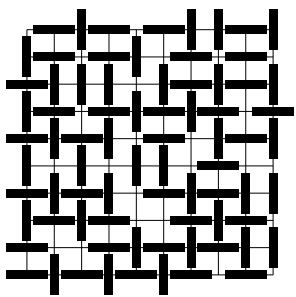,width=4cm,angle=-90}}
        \vspace*{0.1cm}
\caption{Tetris Model: Sketch of the local arrangement of particles in the
Tetris model. The sites of a square lattice can host elongated particles shown 
as rectangles.  The width and length of the particles induce
geometrical frustration. }
\label{Fig1}
\end{figure}

In the simplest version of the model, particles are represented by 
rectangles of uniform size $a\times b$ which are distributed on the sites 
of a square lattice with only two possible orientations (length of the 
particle aligned along one of the principal axes).  
The size of the rectangle (in mesh units) is chosen so
that $a>.5$ and $a+b<1$.  Geometrical frustration results from the
constraint that two particles should not overlap. Fig.~\ref{Fig1}
shows a local arrangement of the particles.

We now consider a system of size $L_x\times L_y$ with periodic
boundary conditions along the horizontal direction 
(parallel to $x$) and the
principal axis of the square lattice pointing along the directions
$(1,\pm 1)$. Gravity is along the $-y$ direction.  This system is filled
with Tetris particles by random deposition under gravity {\it i.e.}
particles having one or the other orientation are 
dropped from a random position at the top. They then fall
downwards to any of the two nearest neighbor sites available and so on
till they reach a site from which they cannot fall any further 
due to hard-core repulsion of other particles present.  This defines
the initial state of the system.

The system is then progressively emptied from the bottom row by removing
one single particle at a time.  Once a particle is removed, other
particles may fall down, and induce what we call an ``internal
avalanche'', i.e. a restructuring of the medium in the bulk to recover a
new stable configuration in which particles can no longer move downwards.
A new particle is then deposited under gravity on top of the system.
This process conserves the number of particles in the system, and after a
large number of avalanches, a steady-state is established
(in which we make all our measurements).
The number of particles which move after the removing of a single grain
is by definition the avalanche size.

From the definition of the model, it appears to be a good candidate for
exhibiting self-organized criticality.  A minimal flow is forced through
the system, since the system is required to relax to a stable configuration
before a new particle is removed from the bottom line.  If the system is
very dense, the progressive removal of particles from the bottom line
will decompactify it. If it is too loose, then avalanches will
propagate easily to the top surface and hence will rapidly increase
the density.  This competition of two effects can be expected 
to lead the system to states such that large avalanches have a 
vanishing but non zero probability to occur.

\begin{figure}
\bigskip
\centerline{
      \psfig{figure=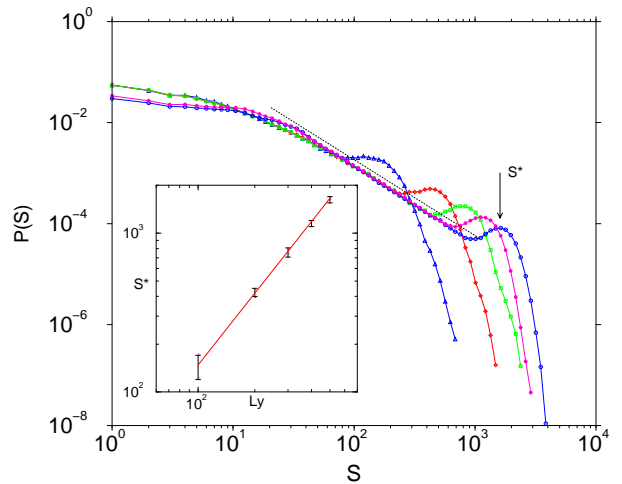,width=8cm,angle=-90}}
        \vspace*{0.1cm}
\caption{Avalanche Distribution:
Log-log plot of the avalanche size distribution obtained in systems of
size $L_x=100$, and $L_y=100,200,300,400,500$. The distribution
shows a scaling region characterized by an exponent $\tau\simeq 1.5$ 
and a  well defined bump for large avalanches. 
The dotted line is a power-law 
fit of exponent $\tau\approx 1.5$.
The top of the bump scales with the height of the system as 
$S^* \sim {L_y}^{1.5}$ (see inset).} 

\label{Fig2}
\end{figure}

Related models have been proposed in the past and it has been
claimed that the avalanche size distribution is power-law
distributed \cite{Ball}. Indeed, such a statistics recorded over 
moderate
systems gives rise to a distribution which can be reasonably fitted
by a power-law as shown in Figure~\ref{Fig2} (see ref.\cite{KriLor} for a
more detailed discussion).


The avalanche size distribution 
exhibits the occurrence of a well defined bump at the maximum
avalanche size (as also observed in other models).
This bump could simply represent a finite-size 
effect due to a looser surface packing.
However we find that the bump is much bigger than what
is expected from just this argument and it is not possible to collapse
curves obtained with different system sizes satisfactorily 
with the usual finite-size scaling. 
The reason is deeply related to the space-time inhomogeneities
self-generated by the dynamics. Extensive numerical simulations 
reveal that the bump displays the following characteristic features:
it corresponds to avalanches of a special type whose typical size 
$s^*$ scales only with the height of the system and is {\em independent} 
of its width.
\be
s^*\approx A L_y^{\alpha}
\ee
We found $\alpha =1.5$ as shown in the inset of Fig. ~\ref{Fig2}.

Since the size of these avalanches is well defined, we could identify
them easily in a time series, and thus study the distribution of time
intervals, $T$, between such avalanches.  
We find that the distribution of $T$ is peaked signifying  
the occurrence of big avalanches at a well-defined frequency.
This also implies a screening effect inhibiting the occurrence 
of large avalanches close to each other.
Therefore, the system displays memory effects over large time intervals
which is one of the features of what we call self-structuring.

\begin{figure}
\bigskip
\centerline{
      \psfig{figure=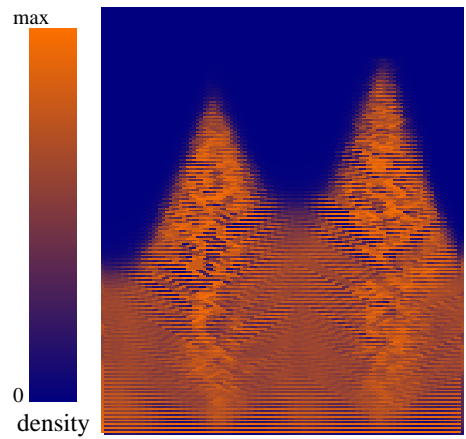,width=6cm,angle=0}}
        \vspace*{0.1cm}
\caption{Density Map:
Color plot of the time-averaged density at every site of the lattice. 
Alternating low (darker) and high (brighter) density 
channels are clearly visible.  Most large
avalanches occur in these channels.}

\label{Fig4}
\end{figure}

The rules of the model are time independent, and thus if a memory effect
emerges, it has to be encoded in the structure of the medium as
spatial inhomogeneities. The peculiar scaling of the large avalanches 
(Eq. 1) also points to the same.
In order to study this point more quantitatively, we studied the
time-average of the local density at every single site (i.e. the
average probability of occupation of a small region centered around 
the site). 
Figure~\ref{Fig4} shows a plot of this time-average density map
on a system of size $L_x=100$, and $L_y=200$ (averaged over $10^{5}$
time steps).  We observe clearly on this map alternating channels of
low and high density
regions which extend from the bottom to the top of the system.  
These channels
have a fixed width $w$ independent of $L_x$ (provided the latter is large
enough) and scaling with $L_y$ as $L_y^{0.5}$ \cite {kudrolli}. 
It is important to note that these channels were not present
initially but have been progressively carved out by the repeated
passages of avalanches.  A qualitative interpretation of this phenomenon
is that avalanches propagate more easily in less dense regions, and
hence regions of higher density are progressively quenched in the medium.
These channels become more fuzzy close to the bottom line because the
continued removal of particles at the bottom 
forces the flux to be uniform along this line.
It is also important to note that the range of variation of the density
is quite moderate (order of $10^{-2}$), and hence it is necessary to
perform a very long time average to be able to capture this effect.  A
snapshot of the system at a single time does not reveal these channels.
This spatial organization is the second hallmark of self-organized
structuring.

\begin{figure}
\bigskip
\centerline{
      \psfig{figure=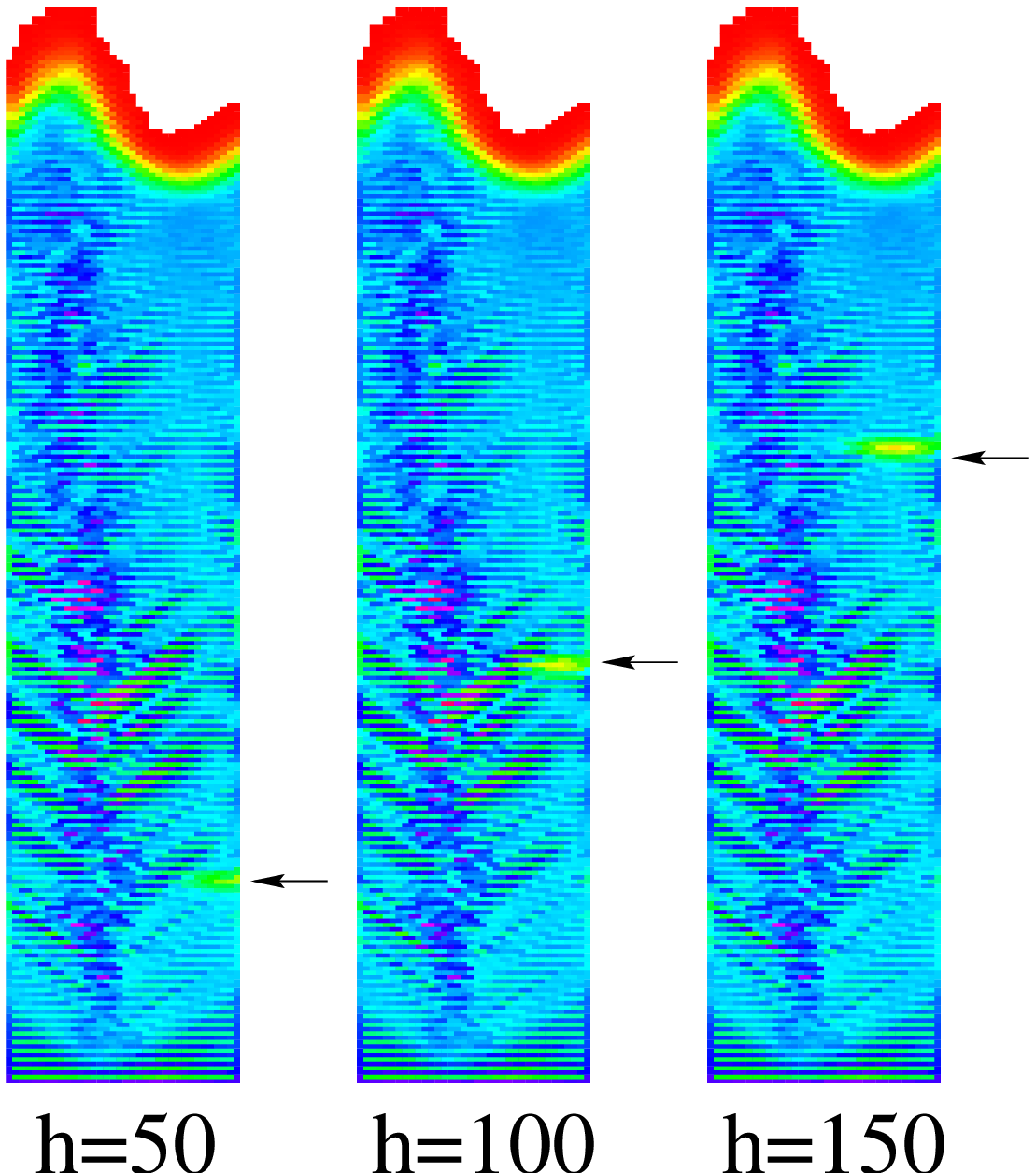,width=8cm,angle=0}}
        \vspace*{0.1cm}
\caption{Bubbles:
Emergent spatial structure in a time average performed over the avalanches
while they are crossing a certain height $h$ marked in the figure. 
The front (marked by the arrows) propagates upwards as a localized low 
density (depicted here by a lighter colour)
region (the ``bubble'') preserving its size and shape.}
\label{Fig5}
\end{figure}

We are now in a position to study the space and time structure of the
large avalanches.  We tailored a system of width $L_x=40$ and $L_y=200$
so that only a single channel would appear.  We repeated the above procedure
of time-averaging the density of particles, but now {\bf during} an
avalanche, when it has reached a given height.  
The resulting maps are shown in
Figure~\ref{Fig5}.  We now see clearly that a large avalanche consists
of a ''bubble'' of low density which propagates along
the previously shown channel.  This bubble is initially rather
diffuse but becomes mature at intermediate heights, and preserves its
shape and size like a solitary wave, as it propagates upwards to the free
surface.  

These results show that large avalanches consist of bubbles propagating
in low density channels at regular time intervals.  The latter is just
the time needed to nucleate such a bubble by the accumulative effect
of small avalanches which deposit voids in the medium.
This nucleation stage is the one where (apparently power-law
distributed) small avalanches are observed. Further the scaling
exponent shown in Eq. 1 can be explained considering the fact that
large avalanches are constrained by the size of the low-density channels 
which have a width vs. height scaling as mentioned above.

 
We have demonstrated that the peculiar statistics of avalanches in the
model discussed above is due to a large scale spatial organization and
long time memory which naturally emerge from complex interactions between
elements.  This model is however just one example of such a phenomenon,
(in the same way as the sandpile model of Ref.\cite{BakTan} was only an
example of self-organized criticality).  We now claim that a similar
scenario may also be at play in other models which have often been 
discussed in the context of self-organized criticality.

One famous example is the Burridge-Knopoff model\cite{BurKno}, a
one-dimensional spring-block model for friction introduced in order to
understand the dynamics of earthquakes.   The model consists of a
one-dimensional chain of blocks connected by springs and driven through
additional soft springs connected to a rigid rod moving at a vanishingly
small velocity.  The friction law which governs the interaction between
the blocks and their support is a velocity weakening law. This model was
studied in details by Carlson and Langer\cite{CarLan,CarLan2} 
who reported that the
statistics of slip events was power-law distributed, a property
reminiscent of the observed Gutenberg-Richter law in earthquakes
statistics\cite{Sho}, and of self-organized criticality.  The most
remarkable aspect of this model is that no noise is explicitly introduced,
and a non uniform motion is only due to an intrinsic instability of the
velocity weakening friction law.  

In spite of the fact that this point did not raise much attention, the
statistical distribution of the length of slip events displays an
initial part (reasonably described by a power-law distribution) 
along with a significant bump at sizes equal to the system size.  Infact, 
in terms of energy released during a slip, it is the latter which
contribute the most (which is somewhat in conflict with geophysical
data).  A detailed numerical investigation\cite{SchVil} allowed one of 
us to get some information about the structure of these large slip events.  
They can be shown to consist of localized pulses propagating through  
the system and occurring at a well-defined time interval.  
Driving the system at a non vanishing speed\cite{SchVil2} gives rise 
to similar pulses occurring at fixed time intervals: an organized 
state where noise disappears.  This point indeed shows that the time 
invariance of this model is broken, and thus that the model belongs 
to the class of self-organized structuring models.  

After the initial proposal of Bak {\it et al}\cite{BakTan} who
exemplified the notion of SOC by a sandpile model, numerous 
attempts have been made to measure power-law distribution of 
avalanches in real granular media.  It was soon realized that 
for large system sizes, the statistics of avalanches was no longer 
a power-law but consisted rather of a peaked distribution
of large avalanches, plus a tail for smaller ones.  This form was
interpreted as resulting from an hysteresis in the repose angle of
granular media rather than a single {\it critical} slope as would be
predicted by a SOC scenario. We can understand this effect from the
point of view of self-structuring, as the consequence
of a slow nucleation effect of avalanches. From simple conservation laws 
this induces a hysteresis in the angle of repose. 


Needless to say, our objective is not to deny the relevance and
importance of self-organized criticality.  On the contrary, the number
of examples where genuine SOC has been demonstrated (e.g. the Abelian
Sandpile model\cite{Dha}) speak for themselves.  
The point we wish to stress is that in some cases, large scale 
instabilities may modulate the response of the system at a 
fixed frequency, giving rise to a statistics of events or avalanches 
which is not purely scale-invariant but rather displays a characteristic size
(which depends on the structures  spontaneously arising in the
steady state and thus not trivially related to the 
system size),  a feature we termed ``self-organized structuring''.
Moreover, the breaking down of the translational time invariance of
the system can be accompanied by a corresponding spatial structure
which emerges from the dynamics of the system.  In contrast, usual SOC is
expected to be observed in similar conditions for stable systems, where
no long-range structures survive, either in space or time.
We also note that the stability or instability of these large structures
might be difficult to analyze a priori, and thus establishing the
self-organized critical or self-organized structuring
character of the system may require large scale numerical 
studies of these models, in order to distinguish these structures from the
the large amplitude small scale background noise inherent to these models.

{\bf Acknowledgements} We acknowledge useful and inspiring 
discussions with H.J. Herrmann, S. S.  Manna, J. Schmittbuhl and
J.P. Vilotte.

\begin{thebibliography}{999}

\bibitem{Mandelbrot} Mandelbrot, B.B.  {\sl ``The Fractal Geometry of Nature} 
Freeman, New York (1982).

\bibitem{BakTan} Bak, P., Tang, C. \& Wiesenfeld, K. 
{\it Phys. Rev. Lett.} {\bf 59}, 381 (1987).



\bibitem{CarLan} Carlson J. \& Langer, J.S.,
{\it Phys. Rev. Lett.} {\bf 62}, 2632 (1989).

\bibitem{CarLan2} Carlson, J.M., Langer \& Shaw, B.E.,
  {\it Rev. Mod. Phys.} {\bf 66}, 657 (1994).

\bibitem{soc-economy} Bak, P., Chen, K., Scheinkmen, J. \& Woodford, M.
{\it Ricerche Economiche} {\bf 47}, 3 (1993).

\bibitem{SneBak} Sneppen, K. \& Bak, P.,  {\it Phys. Rev. Lett.}  
{\bf 71}, 4083 (1987).

\bibitem{grain} For a recent introduction to the overall phenomenology
see Proceedings of the NATO Advanced Study Institute on
 {\it Physics of Dry Granular Media}, Eds.
 Herrmann H.J. {\it et al}, Kluwer Academic Publishers, Netherlands
 (1998).  



\bibitem{KriLor} Krishnamurthy, S., Loreto, V., Herrmann, H.J., Manna, S.S. \&
Roux,  S., to appear in {\it Phys. Rev. Lett.} (1999).


\bibitem{Ball} Snyder, R.E., \& Ball, R.C.,  {\em Phys. Rev. E},  
{\bf 49}, 104 (1994). 



\bibitem{BurKno} Burridge, R. \& Knopoff, L., 
{\it Bull. Seismol. Soc. Am.} {\bf 57}, 3411 (1967).

\bibitem{Sho} Sholtz, C.H. 
{\sl ``The Mechanics of Earthquakes and Faulting''}, 
Cambridge Univ. Press, Cambridge (1990).

\bibitem{oslo} Frette, V., Christensen, K., Malthe-S\o renssen, A.,
Feder, J., J\o ssang, T. \& Meakin, P., {\em Nature} {\bf 379},  49 (1996).


\bibitem{tetris} Caglioti, E., Loreto, V., Herrmann, H.J. \& Nicodemi, M.,
{\it Phys. Rev. Lett.} {\bf 79}, 1575 (1997). Caglioti, E., 
Krishnamurthy, S. \& Loreto, V., {\it Random Tetris Model}, 
unpublished (1999).


\bibitem{Knight} Knight, J.B., Fandrich, C.G., Ning Lau, C., Jaeger,
H.M. \&  Nagel, S.R., {\em Phys. Rev. E} {\bf 51}, 3957 (1995). 

\bibitem{segtet}
Caglioti, E., Coniglio, A., Herrmann, H.J., Loreto V. \& Nicodemi, M.,
{\em Europhys. Lett.} {\bf 43}, 591 (1998).


\bibitem{kudrolli} Curiously an experiment similar in spirit
to the dynamics we propose finds similar results, see
 A. Samadani, A. Pradhan \& A. Kudrolli, {\em cond-mat}/9905303.


\bibitem{SchVil} Schmittbuhl, J., Vilotte, J.P. \& Roux, S., 
{\it J. Geophy. Res.} {\bf 101}, 27741 (1996)

\bibitem{SchVil2} Schmittbuhl, J., Vilotte, J.P. \& Roux, S., 
{\it Europhys. Lett.} {\bf 21}, 375 (1993)

\bibitem{Dha} Dhar, D., {\it Phys. Rev. Lett.} {\bf 64}, 1613 (1990).

\end{thebibliography}
\end{document}